\documentclass{sig-alternate-05-2015}

\usepackage[linesnumbered]{algorithm2e}
\usepackage[caption=false]{subfig}
\usepackage{graphicx}
\usepackage{amsfonts}
 \usepackage{multirow}
 \usepackage{pgfplots}
 \usepgfplotslibrary{external} 
\begin{document}


\numberofauthors{1} \author{
\alignauthor
Brian Brost\textsuperscript{1}~~~~ Yevgeny Seldin\textsuperscript{1}~~~~ Ingemar J. Cox\textsuperscript{1,2}~~~~ Christina Lioma\textsuperscript{1} \\
       \affaddr{ \textsuperscript{1} University of Copenhagen, Department of Computer Science}\\
       \affaddr{ \textsuperscript{2} University College London, Department of Computer Science}\\
       \email{brian.brost@di.ku.dk~~~~ seldin@di.ku.dk~~~~ ingemar@ieee.org~~~~ c.lioma@di.ku.dk}
}

\title{Multi-Dueling Bandits and Their Application to Online Ranker Evaluation} 
\subtitle{Extended Version}

\date{30 July 1999}

\maketitle

\begin{abstract}




New ranking algorithms are continually being developed and refined, necessitating the development of efficient methods for evaluating these rankers. Online ranker evaluation focuses on the challenge of efficiently determining, from implicit user feedback, which ranker out of a finite set of rankers is the best. 

Online ranker evaluation can be modeled by \emph{dueling bandits}, a mathematical model for online learning under limited feedback from pairwise comparisons. Comparisons of pairs of rankers is performed by \emph{interleaving} their result sets and examining which documents users click on. The dueling bandits model addresses the key issue of which pair of rankers to compare at each iteration, thereby providing a solution to the exploration-exploitation trade-off.

Recently, methods for simultaneously comparing more than two rankers have been developed. However, the question of which rankers to compare at each iteration was left open. We address this question by proposing a generalization of the dueling bandits model that uses simultaneous comparisons of an unrestricted number of rankers.

We evaluate our algorithm on synthetic data and several standard large-scale online ranker evaluation datasets. Our experimental results show that the algorithm yields orders of magnitude improvement in performance compared to state-of-the-art dueling bandit algorithms.
\end{abstract}

\section{Introduction}

Generation of ranked lists is a common requirement, particularly in the context of web search, paid placement advertising, and recommender systems. Rankers, i.e., algorithms for generating ranked lists, are continually being developed and refined. Taking advantage of the growing pool of rankers requires efficient evaluation of their relative quality, since time spent on evaluation of suboptimal rankers has to be minimized.

Evaluation of rankers can be done online by presenting the ranked lists produced by rankers to users and then inferring the quality of the rankers by analyzing users' clicks and other forms of behaviour. Online evaluation of rankers has become increasingly popular, partly because user behaviour can be easily logged with no additional effort from the user. This provides online evaluation methods with inexpensive access to large amounts of timely training data \cite{chapelle2012large}. One of the key drawbacks of online evaluation methods is that the outputs of new, potentially poor, rankers need to be presented to actual users. If a new ranker turns out to be poor, then users will be presented with poor results and, in the worst case, might abandon the service \cite{hofmann2011balancing}. Conversely, if new rankers are not presented there is a risk of overlooking better rankers in the pool of rankers. In online learning the question of determining a proper exploration level is known as the \emph{exploration-exploitation tradeoff}. 

In online evaluation, it is usually easier for users to make relative judgements, rather than absolute judgements. For example, it is easier to say that document A is more relevant for a certain query than document B, than to say how relevant it is. Similarly, rankers A and B can be compared by \emph{interleaving} their result lists and examining which documents a user clicks on. Interleaving methods were found to require 1-2 orders of magnitude less data than absolute metrics to detect even small differences in retrieval quality \cite{chapelle2012large}.


When using interleaving to compare pairs of rankers, it is critical to determine which two rankers to interleave at each comparison, i.e. to resolve the exploration-exploitation tradeoff. \emph{Dueling bandits} is an elegant mathematical framework that provides a principled way for dealing with the exploration-exploitation trade-off in learning with relative preference feedback from pairwise comparisons \cite{yue2012k}. It has been successfully applied to online ranker evaluation based on interleaving \cite{zoghi2014relative,zoghi2014relativesampling,zoghi2015mergerucb}.


More recently, interleaving has been generalized to \emph{multileaving} which permits more than two rankers to be compared in a single comparison \cite{schuth2014multileaved, schuth2015probabilistic, brost2016improved}. Prior work on multileaving has shown that repetitive simultaneous comparison of fixed sets of rankers through multileaving is similarly accurate in identifying the best ranker in the set and faster than sequential comparisons of pairs of rankers through interleaving. However, prior work focused only on the comparisons themselves, and did not address the key issue of selecting subsets of rankers for each comparison. This means that all rankers, both good and bad ones, were used in all the comparisons. This simple approach has several disadvantages. Firstly, since poor rankers are participating on all the comparisons the quality of the multileaved lists throughout the evaluation process is poor. And secondly, very poor rankers that could potentially be eliminated early in the process continue being evaluated, which does not allow the comparisons to focus on rankers whose quality is harder to distinguish. 


 

We extend the dueling bandit framework and propose a \emph{Multi-Dueling Bandit} algorithm that provides an intelligent selection of rankers for simultaneous comparisons and improves the trade-off between exploration and exploitation. Our experimental evaluations of the new algorithm on both synthetic and real web search learning-to-rank datasets show that our algorithm yields orders of magnitude improvement in performance compared to state-of-the-art dueling bandit algorithms.

Section~\ref{sec:relatedWork} provides background material and reviews prior work. Section~\ref{sec:problem} formalises the problem of learning with relative comparisons between multiple options as a $K$-armed \emph{multi-dueling bandit} problem. Section~\ref{sec:algorithm} describes our proposed algorithm for solving this problem. Section~\ref{sec:experimentalevaluation} presents the experimental evaluation. This is an extended version of the paper published at CIKM \cite{brost2016multishort}.


\section{Background and Related Work}
\label{sec:relatedWork}

Our work builds on two main bodies of literature. The first is concerned with \emph{selecting} promising rankers from a pool of rankers and the second is concerned with \emph{combining} their ranked lists and \emph{evaluating} their relative quality. In previous work, {\em selecting} rankers has been framed as a dueling bandit problem, and {\em combining} and {\em evaluating} the selected rankers have been done via interleaving and multileaving. The corresponding prior work is discussed next.

\subsection{Dueling Bandits}

The simplest approach to selection of rankers is to select one ranker at a time. This approach can be modeled within the multiarmed bandit framework \cite{auer2002finite}, where we associate rankers with arms. At each round of a multiarmed bandit game the player picks one of $K$ possible arms (in our case rankers) and observes the reward of that arm. Rewards of arms that were not selected remain unobserved. 
The goal is to select arms, so that the cumulative regret, defined as the difference between the reward of the unknown best ranker and the reward of the chosen ranker, is minimized.

An alternative approach is to select pairs of rankers, present some combination of their ranked lists, and infer their relative quality from user clicks (technical details of how this can be done are discussed below). 
It has been shown that relative comparisons require 1-2 orders of magnitude fewer comparisons to determine quality differences between rankers than single-arm payouts \cite{chapelle2012large}. Learning from limited relative feedback from pairwise comparisons can be modeled as a $K$-armed \emph{dueling} bandit problem \cite{yue2012k}. The goal is to select pairs of rankers, so that a mix of their rankings will be almost as good as the ranking of the best ranker in the pool. (We note that the model permits selecting the same ranker twice, as the first and the second element of the pair.) There are several possible definitions of the best ranker for pairwise comparisons. The most common definition is the \emph{Condorcet winner}, which is a ranker that is (pairwise) better that any other ranker in the pool. Note that a Condorcet winner is not guaranteed to exist (imagine a situation with three rankers A, B, and C, where A is better than B, B is better than C, and C is better than A). Nevertheless, many dueling bandit algorithms assume the existence of a Condorcet winner and we follow suit.


The $K$-armed dueling bandit problem was introduced by Yue \emph{et al.} \cite{yue2012k}, who also presented an algorithm called \emph{interleaved filtering} (IF) for solving it. In IF, arms are eliminated sequentially by comparing them with the best currently known arm until they are defeated with sufficient confidence. 
Yue \emph{et al.} \cite{yue2011beat} subsequently proposed an improved algorithm called \emph{beat the mean} (BTM). This algorithm attempted to reduce the number of comparisons needed by focusing on comparing the arm that had been used in the least number of comparisons with a randomly sampled arm that had not yet been eliminated. Yue \emph{et al.} also presented experimental evidence confirming the superior performance of BTM over IF.

Zoghi \emph{et al.} \cite{zoghi2014relative} proposed an algorithm for the dueling bandit setting based on the idea of \emph{relative upper confidence bounds} (RUCB). The algorithm maintains a relative upper confidence bound on the probability that a given arm $i$ is better than another arm $j$. The algorithm then selects an arm $i$ that might be the best, based on its upper confidence bounds relative to all other arms, and then selects the challenger with the highest upper confidence bound relative to $i$. This approach was shown to outperform both IF and BTM. Zoghi \emph{et al.} \cite{zoghi2015mergerucb} subsequently proposed a divide-and-conquer algorithm, \emph{MergeRUCB}, extending their earlier work in \cite{zoghi2014relative}. MergeRUCB was designed to perform well for problems involving a large number of arms. Extensive experiments by Zoghi \emph{et al.} suggest that it substantially outperforms IF and BTM, and that for large numbers of arms it outperforms RUCB too.

Komiyama \emph{et al.} \cite{komiyama2015regret} proposed an algorithm, \emph{Relative Minimum Empirical Divergence} (RMED), which draws arms based on whether an arm has not been compared with other arms sufficiently often, or if it is not substantially beaten by many other arms. To decide if an arm has been sufficiently explored it uses bounds based on the KL-divergence. They showed that this algorithm outperformed RUCB and MergeRUCB. To the best of our knowledge this is currently the best performing dueling bandit algorithm.

All algorithms listed above are limited to pairwise comparisons, whereas our proposed algorithm is based on simultaneous comparisons of multiple items.

\subsection{Comparison Methods}
\label{sec:comp}

Combining ranked lists and evaluating chosen rankers is based on click-through information obtained by \emph{interleaving} (for two rankers) or \emph{multileaving} (for two or more rankers) methods. A variety of interleaving methods have been proposed based on various ways to create the interleaved list and assign credit to rankers, see \cite{hofmann2013fidelity,chapelle2012large} for an overview. 

More recently, interleaving has been generalised to simultaneous comparison of more than two rankers. The proposed multileaving methods include Team Draft Multileave (TDM) \cite{schuth2014multileaved}, Optimised Multileave (OM) \cite{schuth2014multileaved}, Probabilistic Multileave (PM) \cite{schuth2015probabilistic}, and Sample Only Scored Multileave (SOSM) \cite{brost2016improved}. SOSM was found to outperform the other multileaving methods, which can either be less accurate, or fail to scale well with the number of rankers \cite{brost2016improved}. 
We used SOSM to multileave rankers for our multi-dueling bandit algorithm.

\section{The Multi-Dueling Bandit Problem}
\label{sec:problem}

In multi-dueling bandits, at each iteration, $t$, an algorithm selects a subset, $S_t$, of $K$ arms and observes outcomes of noisy pairwise comparisons (duels) between all pairs of arms in $S_t$. In the ranking scenario this corresponds to multileaving the ranked lists of the subset, $S_t$, of rankers and then inferring the relative quality of the lists (and the corresponding rankers) from user clicks. When the size of $S_t$ is limited to 2 the problem reduces to standard dueling bandits. 

Let $P = [p_{ij}]$ be a matrix of probabilities that arm $i$ wins in a pairwise comparison with arm $j$ (it satisfies $p_{ij} = 1-p_{ji}$ and we define $p_{ii}=\frac{1}{2}$). As we have already mentioned, in pairwise comparisons the best arm is not always well-defined (recall the example with A being better than B, B better than C, and C better than A). We follow the assumption in most dueling bandit literature and assume that there exists a Condorcet winner, which is a unique arm $*$ satisfying $p_{*j} > \frac{1}{2}$ for all $j \neq *$. That is, the Condorcet winner $*$ is pairwise better than any other arm $j$. The quality of all arms is then defined by their \emph{regret}, $r(j) = p_{*j} - \frac{1}{2}$, which is a shifted probability of losing to the best arm (this definition also coincides with dueling bandits). Smaller regret corresponds to better quality and the regret of playing the best arm is zero. The quality of a set of arms $S_t$ is defined by the average quality of the constituent arms (the average regret)
\begin{equation}
r(S_t) = \frac{\sum_{j \in S_t}p_{*j}}{|S_t|}-\frac{1}{2}.
\label{eq:regret2}
\end{equation}
The goal of a multi-dueling bandit algorithm is to select subsets of arms $S_1, S_2,\dots$, so that the cumulative regret $\sum_{t=1}^T r(S_t)$ is minimized. All arms have to be selected a small number of times in order to be explored, but the goal of the algorithm is to minimize the number of times when suboptimal arms are selected. On average, simultaneous exploration has lower regret than sequential comparison. 

Simultaneous comparison of more than two arms may affect their pairwise winning probabilities. For example, in ranking, the effective length of a multileaved ranked list is typically limited by 10 items, since users rarely go beyond the first page of results. Therefore, the simultaneous comparison of more than 10 rankers means that some rankers may be compared based on a merged list that does not include their top suggestions. This may affect the estimates of their relative quality. This effect, which we refer to as {\em distortion} may also occur when less than 10 rankers are compared, since the limited length of the merged list does not allow perfect representation of every ranker. The exact level of distortion depends on the data, ranker, and method used for multileaving. The level of distortion of estimates of the pairwise winning probabilities made by SOSM, which was used in our experiments, is evaluated in Section~\ref{sec:expConsistency}. It is important to emphasize that in all our experimental comparisons, except one pathological case, the advantage of parallel exploration outweighed the disadvantage due to distortion in estimates.

\subsection{Multi-Dueling Bandit Algorithm}
\label{sec:algorithm}

The proposed multi-dueling bandit algorithm is based on the principle of ``optimism in the face of uncertainty'' used in many other bandit algorithms. It maintains optimistic estimates of pairwise winning probabilities $p_{ij}$ and plays arms that, according to these optimistic estimates, have a chance of being the Condorcet winner. When there is a single candidate, the algorithm \emph{exploits} this knowledge and plays only that candidate. When there are multiple candidates the algorithm \emph{explores} by comparing them all. We increase parallel exploration by adding additional arms to such comparisons, as described below. 

Our estimates of pairwise winning probabilities are based on empirical counts of wins/losses.
 In order for these estimates to be meaningful the algorithm has to assume that pairwise winning probabilities are \emph{consistent} with the pairwise winning probability matrix $P$, irrespective of the composition of the set $S_t$ (meaning that they are not distorted). More precisely, since correct identification of the Condorcet winner depends on correct estimation of the probabilities $p_{*j}$, it is important that they remain at a certain margin above $\frac{1}{2}$ irrespective of the composition of $S_t$. Incorrect estimation of $p_{ij}$-s for $i,j\neq*$ does not influence identification of the Condorcet winner and, therefore, their distortion does not disturb the operation of the algorithm.

We now describe our algorithm, which is provided in the Algorithm~\ref{alg:RUCBBattling} box. We denote by $n_{ij}(t)$ the number of times up to round $t$ that $i$ and $j$ were compared with each other. Let $w_{ij}(t)$ denote the number of times when arm $i$ beat arm $j$. We break ties randomly, so that $n_{ij}(t) = w_{ij}(t) + w_{ji}(t)$. We compute upper confidence bounds $u_{ij}(t)$ on the probabilities $p_{ij}$:
\begin{equation}
u_{ij}(t)=\frac{w_{ij}(t)}{n_{ij}(t)}+\sqrt{\frac{\alpha \ln t}{n_{ij}(t)}}
\label{eqn:upperBound}
\end{equation}
($u_{ij}$-s are the optimistic estimates of $p_{ij}$-s and they are analogous to those used in \cite{zoghi2014relative} for pairwise comparisons). The first term in $u_{ij}(t)$ is an empirical estimate of $p_{ij}$ and the second term bounds the fluctuations of this estimate with high probability, see \cite{auer2002finite,zoghi2014relative}. The $\alpha$ parameter in the second term controls the width of the upper confidence bound. 


Additionally, we maintain a second wider upper bound $v_{ij}(t)$, which we use to increase parallel exploration. We define $v_{ij}(t)$ by
\begin{equation}
v_{ij}(t) = \frac{w_{ij}(t)}{n_{ij}(t)}+\sqrt{\frac{\beta \alpha \ln t}{n_{ij}(t)}},
\end{equation}
where the parameter $\beta \geq 1$ controls how much wider it is than the upper confidence bound of Equation~\ref{eqn:upperBound}. When there is more than one candidate for a Condorcet winner according to the ``narrow'' confidence bounds in Equation~\ref{eqn:upperBound} an exploration round is triggered and arms that could be Condorcet winner candidates according to the ``wide'' confidence bounds are compared. This leads to some arms being explored preemptively and decreases the overall number of exploration rounds by increasing parallel exploration. 


Given $K$ arms, we define $U_i(t) = \min_{j \in K, j \neq i} \left \{u_{ij}(t)\right\}$, i.e. $U_i(t)$ is the smallest upper confidence bound of $i$. Let $E$ denote the set of potential Condorcet winners, which contains all arms $i$ for which $U_i(t) \geq 1/2$. Additionally, we define $V_i(t) = \min_{j \in K, j \neq i} \left \{v_{ij}(t) \right \}$ and $F$ to be the set of potential Condorcet winners according to the wider upper bounds, that is, all arms for which $V_i(t) \geq 1/2$.


At each iteration of Algorithm~\ref{alg:RUCBBattling}, if there is only a single potential Condorcet winner in $E$, we choose this arm. If there are several potential Condorcet winners, we select all arms in the larger set $F$. In the unlikely event that there are no potential Condorcet winners, we select all arms. The selected arms are compared against each other using multileaving and pairwise wins between the rankers are inferred from the scores produced by the multileaving method. 




\begin{algorithm}
  $W=[w_{ij}] := 0_{K \times K}$ \\
	Play all arms and update the corresponding entries in $W$ \\
 \For{$t=2,\ldots,T$}{
  $U:=[u_{ij}(t)] = \frac{w_{ij}(t)}{n_{ij}(t)}+\sqrt{\frac{\alpha \ln t}{n_{ij}(t)}}$, $u_{ii}(t)=1/2$ \\
  $V:=[v_{ij}(t)] = \frac{w_{ij}(t)}{n_{ij}(t)}+\sqrt{\frac{\beta \alpha \ln t}{n_{ij}(t)}}$, $v_{ii}(t)=1/2$ \\
  E = \{$i$ s.t. $U_i(t) \geq 1/2$\}  (The set of potential champions according to $U$) \\
  F = \{$i$ s.t. $V_i(t) \geq 1/2$\}  (The set of potential champions according to $V$) \\
  \uIf{$|E| > 1$}{
  Choose all arms $f \in F$ for comparison and update the corresponding entries in $W$ \\ 
  } \uElseIf{$|E|=1$}{
  Choose the arm $e \in E$ \\
  } \uElse{
  Choose all arms for comparison and update the corresponding entries in $W$ \\
  }
 }
 \caption{Multi-Dueling Bandit (MDB) Algorithm.}
 \label{alg:RUCBBattling}
\end{algorithm}

\section{Experimental Evaluation} 
\label{sec:experimentalevaluation}

We next present the experimental evaluation of our Multi-Dueling Bandits (MDB) algorithm. 

\subsection{Experimental Setup}

We begin by describing our experimental setup. 

\subsubsection{Baselines}

We compare our MDB algorithm to three state-of-the-art dueling bandit algorithms, namely RUCB and MergeRUCB, both implemented in the freely available software package Lerot \cite{schuth2013lerot}, and RMED1 \cite{komiyama2015regret}. As per \cite{zoghi2015mergerucb}, we set the $\alpha$ parameter of Equation~\ref{eqn:upperBound} for RUCB to 0.51, and to 1.01 for MergeRUCB. For RMED1 we use the same parameter setting as \cite{komiyama2015regret}: $f(K)=0.3K^{1.01}$. To select the parameters for MDB, we carried out a grid search on the grid $\{ 0.5, 1, 1.5\} \times \{ 1.25, 1.5, 2, 4\}$ on a separate dataset, specifically the validation set of the YLR1 dataset, and found the best parameters to be $\alpha=0.5$ and $\beta=1.5$. We used these as our parameter settings for MDB for all other experiments. 


\subsubsection{Datasets}

We first compare the algorithms on artificial datasets where each arm has a utility which defines its winning probability against other arms, similar to the experiments proposed in \cite{ailon2014reducing}. In each iteration, for the arms chosen by the dueling or multi-dueling bandit algorithm, we sample from normal distributions with mean given by the utility of the arms, and unit variance to obtain scores for each arm. The arm utilities used are listed in Table~\ref{tab:util}. They were chosen to provide problem instances where the quality of the best arm was progressively less distinct from that of the other arms, and where the impact of increasing the number of arms could be isolated. 

\begin{table*}
\caption{Datasets used for artificial utility based experiments.}
\begin{center}
\begin{tabular}{ | l || c |  }
  \hline
  Dataset & Distributions of Utilities of arms   \\ \hline
  1good5poor & 1 arm with utility $0.8$, 5 arms with utility $0.2$  \\
  1good50poor & 1 arm with utility $0.8$, 50 arms with utility $0.2$ \\
  1good200poor & 1 arm with utility $0.8$, 200 arms with utility $0.2$ \\
  2good4poor & 1 arm with utility $0.8$, 1 arm with utility $0.7$, 4 arms with utility $0.2$  \\
  11good40poor & 1 arm with utility $0.8$, 10 arms with utility $0.7$, 40 arms with utility $0.2$ \\
  41good160poor & 1 arm with utility $0.8$, 40 arms with utility $0.7$, 160 arms with utility $0.2$ \\
  3good3poor & 1 arm with utility $0.8$, 2 arm with utility $0.7$, 3 arms with utility $0.2$  \\
  21good30poor & 1 arm with utility $0.8$, 20 arms with utility $0.7$, 30 arms with utility $0.2$ \\
  81good120poor & 1 arm with utility $0.8$, 80 arms with utility $0.7$, 120 arms with utility $0.2$ \\
  arith6 & 1 arm with utility $0.8$, 5 arms with utilities forming arithmetic sequence between 0.7 and 0.2  \\
  arith51 & 1 arm with utility $0.8$, 50 arms with utilities forming arithmetic sequence between 0.7 and 0.2 \\
  arith201 & 1 arm with utility $0.8$, 200 arms with utilities forming arithmetic sequence between 0.7 and 0.2 \\
  geom6 & 1 arm with utility $0.8$, 5 arms with utilities forming geometric sequence between 0.7 and 0.2  \\
  geom51 & 1 arm with utility $0.8$, 50 arms with utilities forming geometric sequence between 0.7 and 0.2 \\
  geom201 & 1 arm with utility $0.8$, 200 arms with utilities forming geometric sequence between 0.7 and 0.2  \\
  \hline
\end{tabular}
\label{tab:util}
\end{center}
\end{table*}

We also compare the algorithms on four large-scale evaluation datasets summarised in Table~\ref{Table:Datasets}\footnote{Only 519 features are non-zero for YLR Set 1 and only 596 features are non-zero for YLR Set 2. The remaining features are zero for all query-document pairs.}. Since there was no Condorcet winner for the Yandex dataset, we randomly sampled subsets of 200 rankers from the Yandex dataset, selecting the first subset with a Condorcet winner. This subset was used in the experiments involving the Yandex dataset, except those described in Section~\ref{ss:noCondorcet}, where we investigate the behaviour of the algorithms in the absence of a Condorcet winner. 



\begin{table}
\caption{Datasets. Each dataset consists of a number of query-document pairs, together with a relevance judgement for the pair. Each document is represented by a feature vector.} 
\begin{center}
\begin{tabular}{ | l || r | r | r |  }
  \hline
  Datasets & Queries & URLs & Features  \\ 
  \hline
  \hline
  MSLR-WEB30K \footnote{Microsoft Learning to Rank Datasets - http://research.microsoft.com/en-us/projects/mslr/} & 31,531 & 3,771,125 & 136   \\ 
  YLR Set 1 \cite{chapelle2011yahoo}& 19,944 & 473,134 & 700 \\
  YLR Set 2 \cite{chapelle2011yahoo}& 1,266 & 34,815 & 700 \\ 
  Yandex  \footnote{Yandex Internet Mathematics Dataset - http://imat2009.yandex.ru/en/datasets} & 9,124 & 97,290 & 245 \\
  \hline
\end{tabular}
\label{Table:Datasets}
\end{center}
\end{table}

\subsubsection{Ranker Construction}

The datasets and the corresponding rankers form our dueling or multi-dueling bandit problem instances. 
Following \cite{zoghi2015mergerucb}, for each dataset we choose the rankers to be the features of the dataset. That is, for a given feature, we construct a ranker which ranks documents only according to the score of that feature. An example of a feature is the BM25 score of the body of the document, or a document's PageRank. As noted in \cite{zoghi2015mergerucb} this is a somewhat artificial setup from a learning-to-rank perspective, since we are generally interested in comparing different retrieval algorithms using all the features of the dataset, rather than finding the
best individual feature. The benefit of this approach is that it makes the experiments easy to replicate. Furthermore, from the point of view of
\emph{evaluating} dueling and multi-dueling bandit algorithms, the difficulty of a problem instance is affected by the relative performance of the rankers, not their absolute performance. Using the feature rankers is therefore useful for assessing the performance of dueling and multi-dueling bandit algorithms since many of the features perform similarly and are therefore difficult to distinguish using interleaved or multileaved comparisons.

\subsubsection{Simulated User Model}

All experiments, except those using the artificial datasets described in Table~\ref{tab:util}, are conducted using a simulated user model. For each iteration we randomly sample with replacement one query from the pool of queries of the dataset. The dueling or multi-dueling bandit algorithms choose rankers, whose results are then interleaved or multileaved respectively, and presented to a simulated user. 
For the dueling bandit algorithms, we compare pairs of rankers using probabilistic interleaving \cite{hofmann2011probabilistic}, which is the best performing interleaving method to the best of our knowledge. For MDB, we use SOSM, which is the best performing multileaving method to the best of our knowledge \cite{brost2016improved}. Both probabilistic interleaving and SOSM only present the top-10 documents to users. This limit was chosen since it is rare for users to look past the first page of results when using search engines.
Clicks are then generated from a probabilistic user model \cite{hofmann2013fidelity}. The interleaving or multileaving algorithm scores the chosen rankers using the clicks generated by the user model. 



The click model used for these experiments was the navigational user model from \cite{hofmann2013fidelity}, unless otherwise stated. This click model describes a user who inspects the retrieved list of documents from top to bottom, and is more likely to click on a document if it is more relevant, but may interrupt their session with a certain probability, rather than inspect the entire list of retrieved documents. This click model has been used as a standard click model for dueling bandit algorithm evaluation in \cite{zoghi2015mergerucb}.



\subsection{Results}
\label{ss:experimentalresults}

Below we summarize the experimental results for the various experiments. For all figures, the error bars show the standard deviation of cumulative regret across runs for each algorithm at the given time step. 


\subsubsection{Experiments on synthetic data}
\label{ss:synthetic}


We begin by examining how the cumulative regret increases at each iteration for each of the four algorithms. We use synthetic data for two reasons. First, synthetic data does not require interleaving or multileaving. This is because, at each iteration, after selecting the rankers to be compared, the comparison is performed based on drawing random numbers from a normal distribution with mean given by each arm's utility and unit variance. Thus, the performance of each of the four bandit algorithms is independent of the interleaving/multileaving, and only due to the bandit algorithm.
The second reason for using synthetic data is that it allows us to control the relative performance of the individual arms. Clearly, if the best arm is much better that the other arm, the problem is easier than the case when the best arm is only slightly better than other arms. 

\begin{figure*}
  \centering
  \subfloat[][1good5poor]{
  \includegraphics[width=0.275\textwidth]{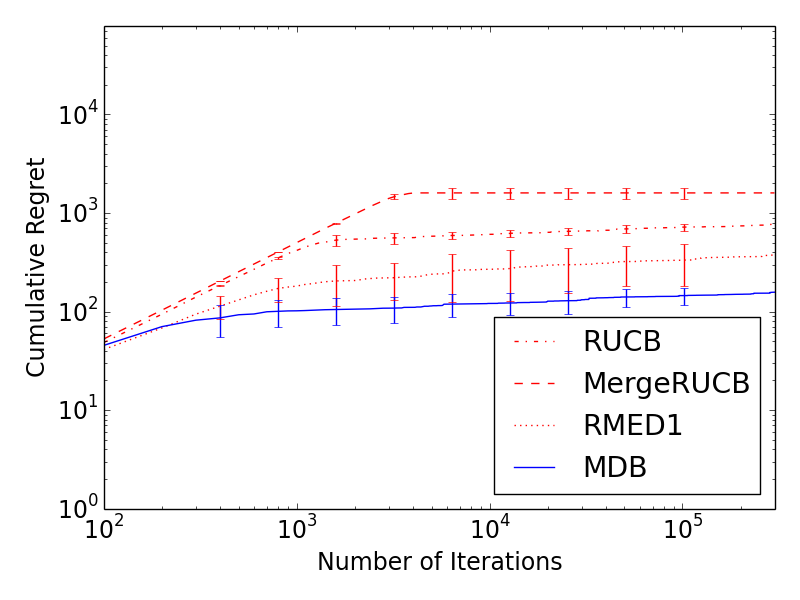}  
  }
  \subfloat[][1good50poor]{
    \includegraphics[width=0.275\textwidth]{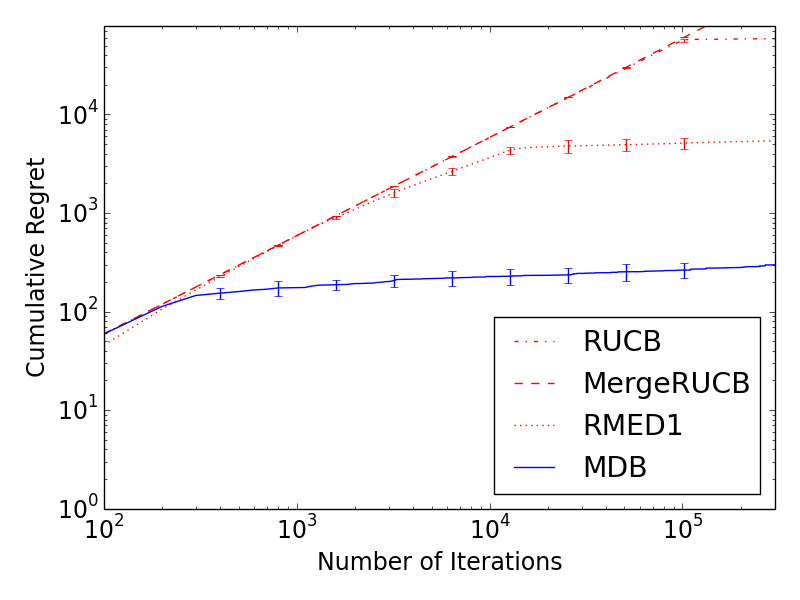}  
  }
  \subfloat[][1good200poor]{
    \includegraphics[width=0.275\textwidth]{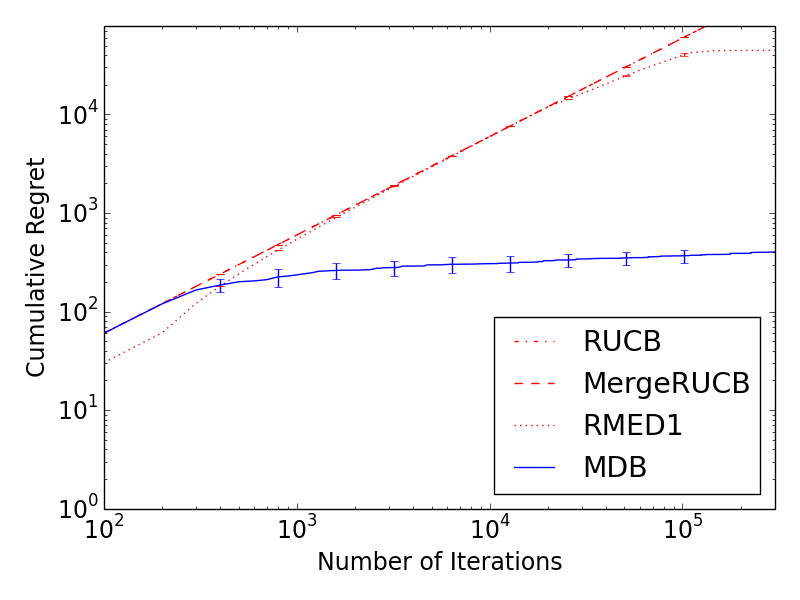}  
  }\\
  \subfloat[][2good4poor]{
  \includegraphics[width=0.275\textwidth]{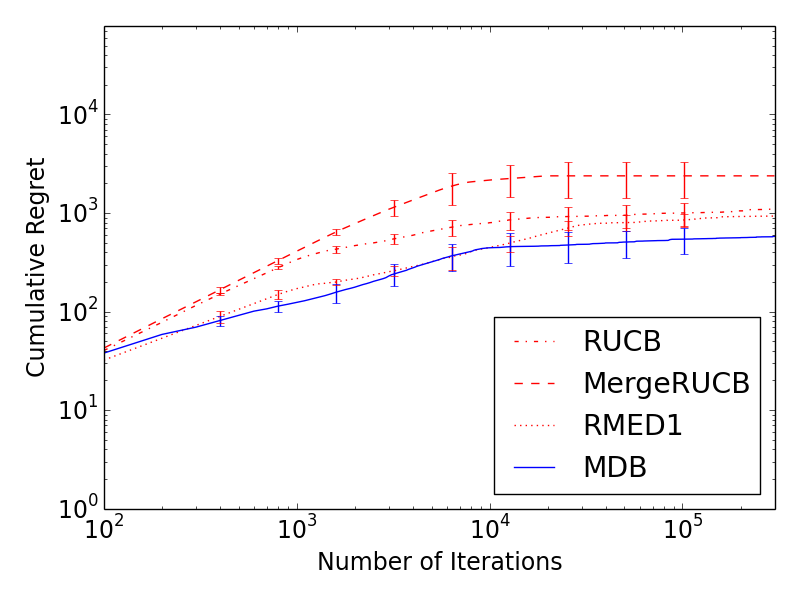}  
  }
  \subfloat[][11good40poor]{
    \includegraphics[width=0.275\textwidth]{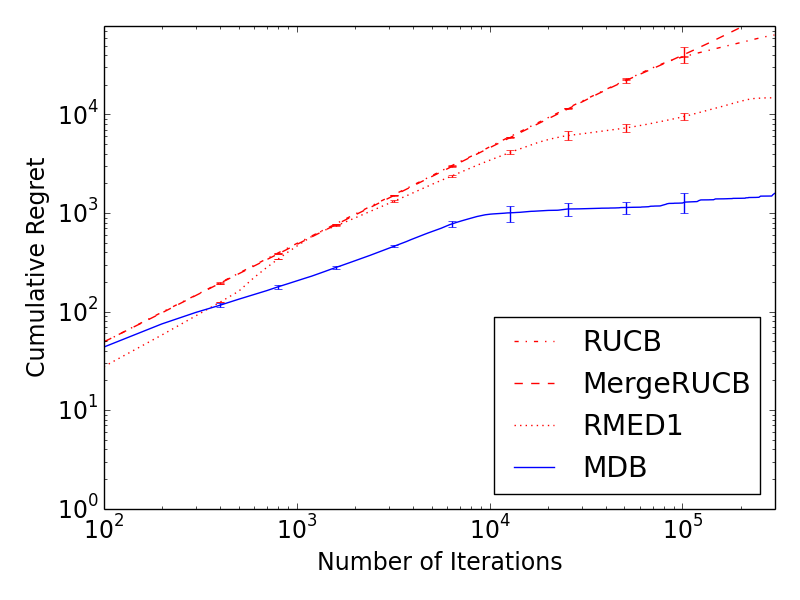}  
  }
  \subfloat[][41good160poor]{
    \includegraphics[width=0.275\textwidth]{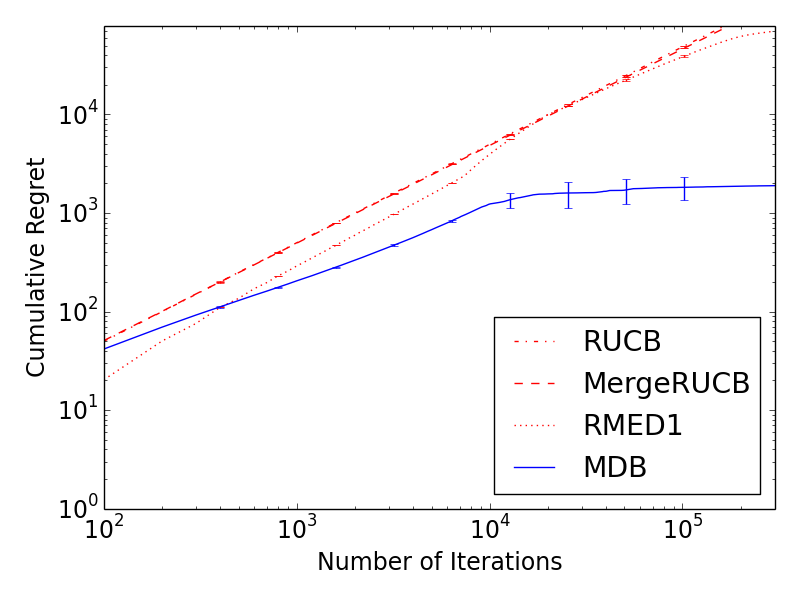}  
  }\\
  \subfloat[][3good3poor]{
  \includegraphics[width=0.275\textwidth]{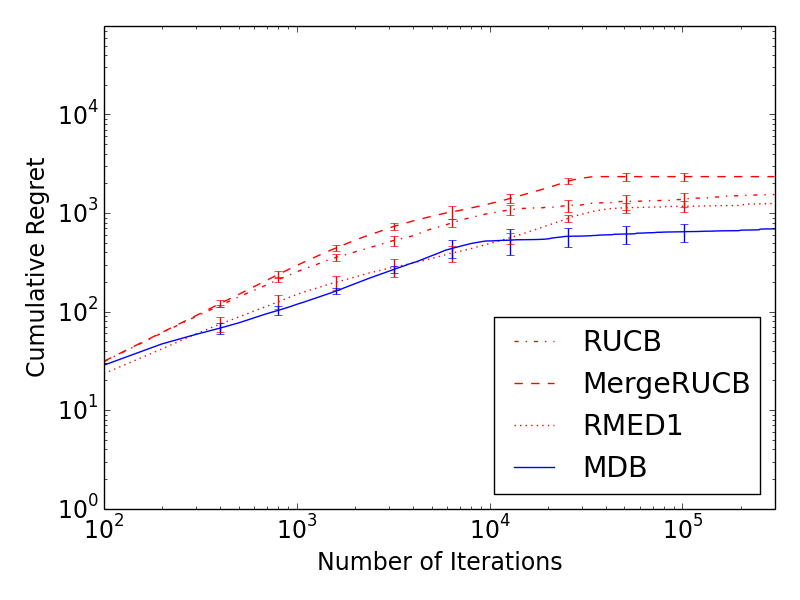}  
  }
  \subfloat[][21good30poor]{
    \includegraphics[width=0.275\textwidth]{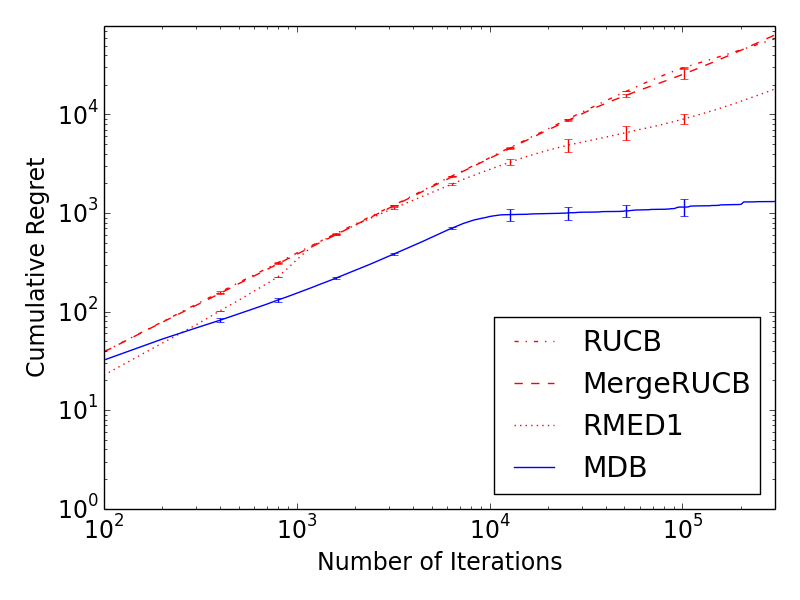}  
  }
  \subfloat[][81good120poor]{
    \includegraphics[width=0.275\textwidth]{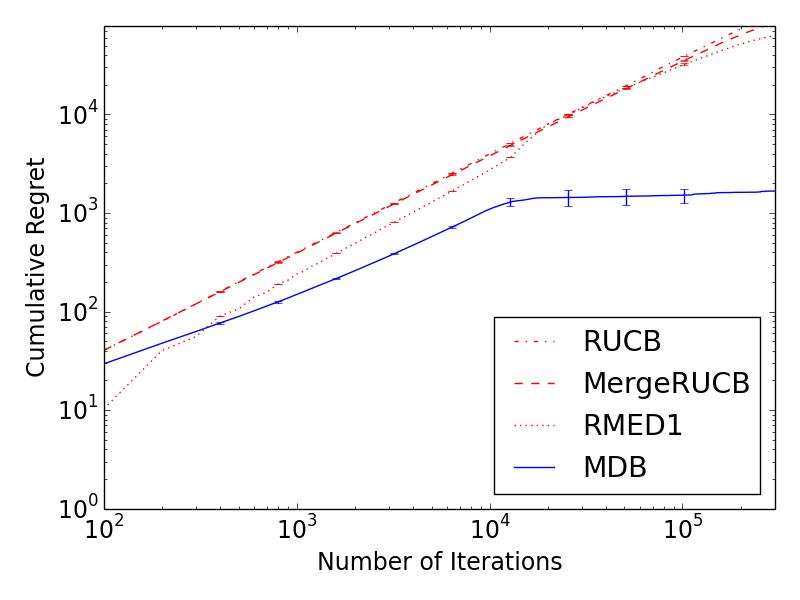}  
  }\\
  \subfloat[][arith6]{
  \includegraphics[width=0.275\textwidth]{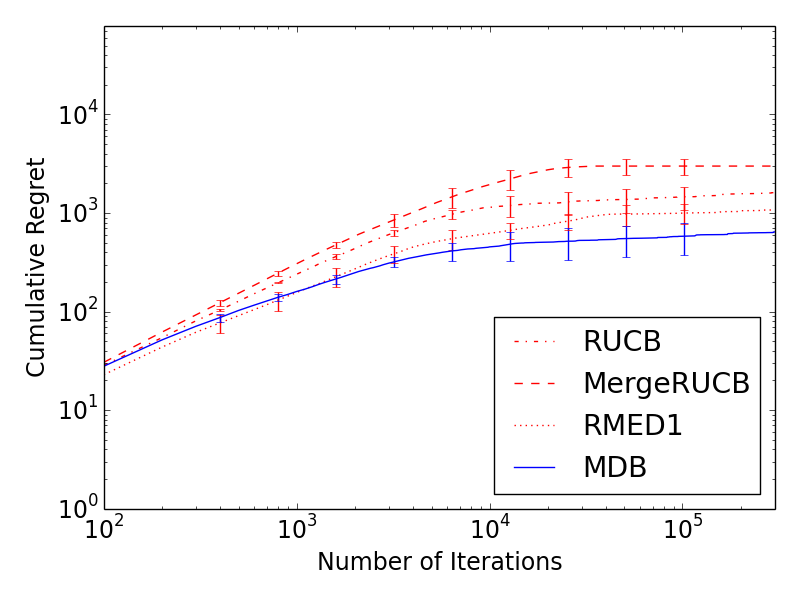}  
  }
  \subfloat[][arith51]{
    \includegraphics[width=0.275\textwidth]{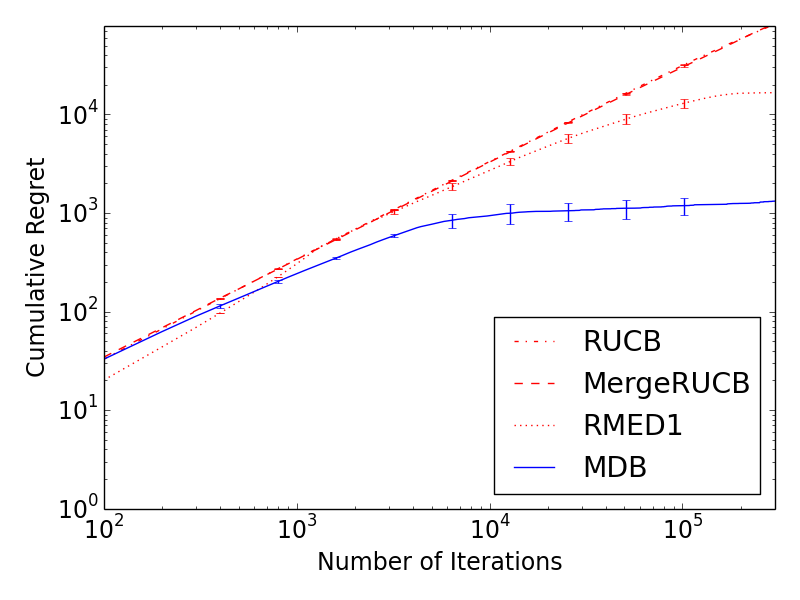}  
  }
  \subfloat[][arith201]{
    \includegraphics[width=0.275\textwidth]{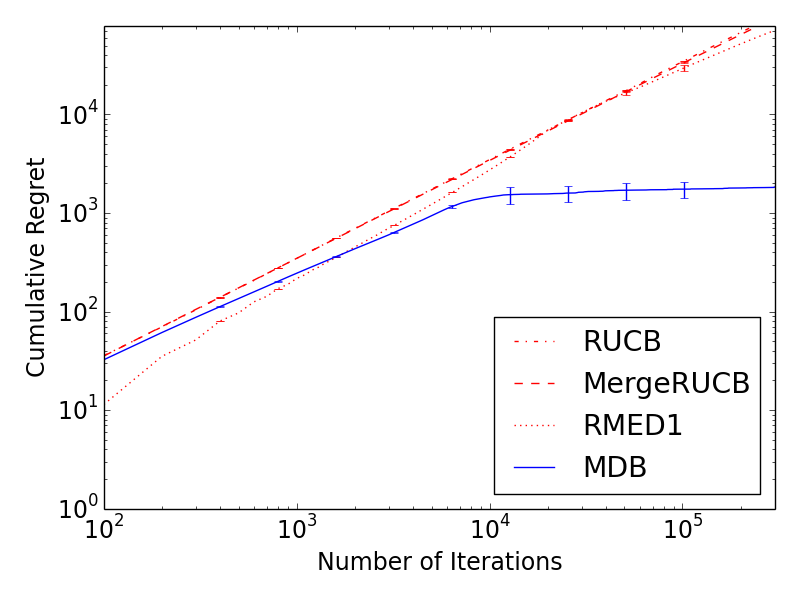}  
  }\\
  \subfloat[][geom6]{
  \includegraphics[width=0.275\textwidth]{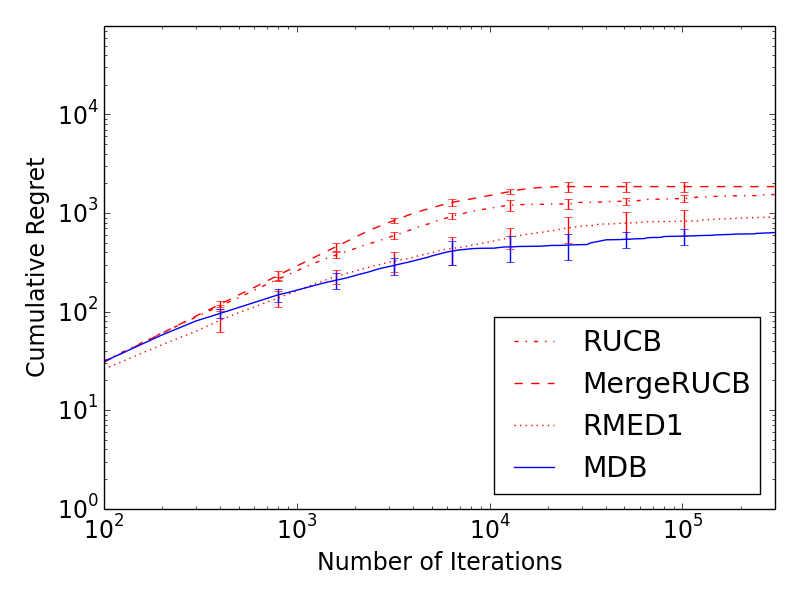}  
  }
  \subfloat[][geom51]{
    \includegraphics[width=0.275\textwidth]{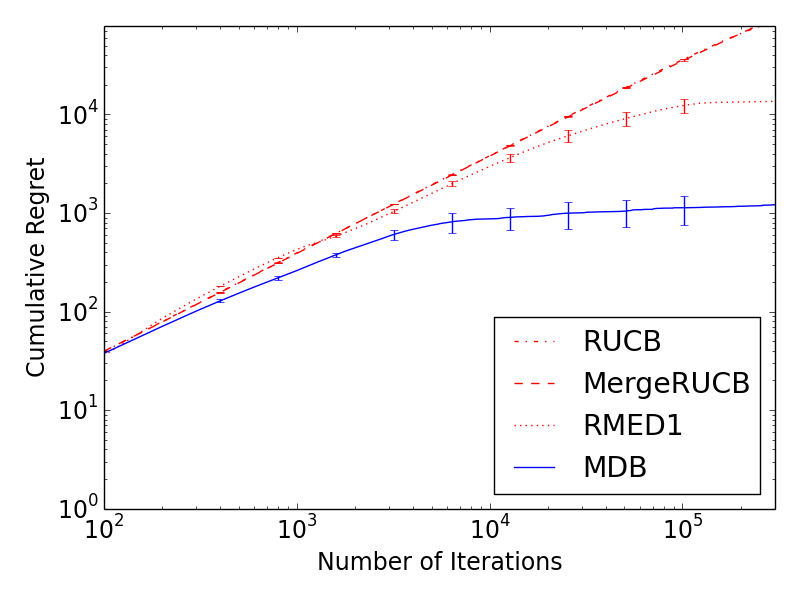}  
  }
  \subfloat[][geom201]{
    \includegraphics[width=0.275\textwidth]{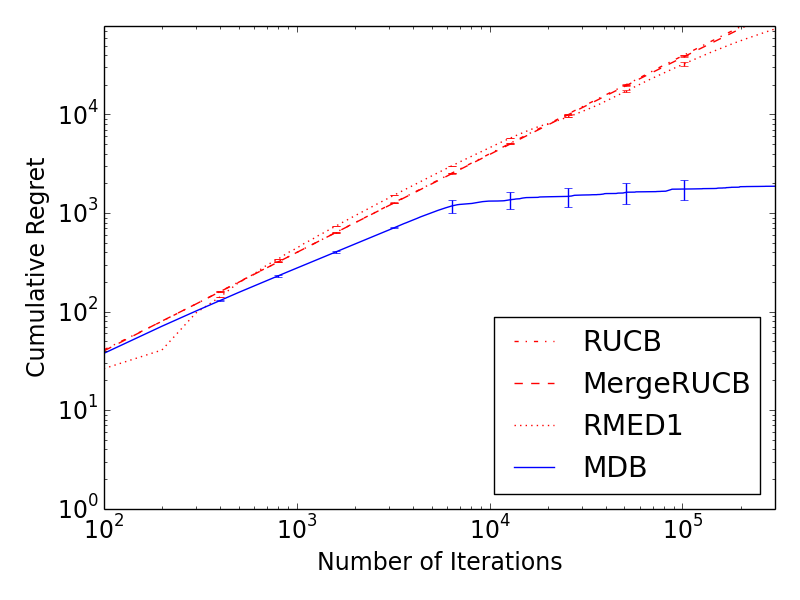}  
  }\\
\caption{Cumulative regret averaged over 10 runs against number of iterations for the 4 algorithms on the datasets listed in Table~\ref{tab:util} 
}
\label{fig:artificial}
\end{figure*}

Figure~\ref{fig:artificial} shows the average cumulative regret against the number of iterations for the 4 algorithms on each of the artificial datasets from Table~\ref{tab:util}. MDB performs better than all the benchmark algorithms for all the datasets. The first column shows the datasets with 6 arms, the second column shows the datasets with 51 arms and the third column shows the datasets with 201 arms. We see that while the regret does not increase noticeably for MDB as we increase the number of arms, the regret of all the dueling bandit algorithms increases substantially. For all the datasets with 51 or more rankers, MDB incurs at least an order of magnitude less regret than the best dueling bandit algorithm, RMED1. 




The results demonstrate that as we increase the number of arms being compared, the advantages of MDB become larger. This advantage is at its most extreme when there is only one good arm, and all other arms are weaker, as in row 1 of Figure~\ref{fig:artificial}. In this case dueling bandit algorithms have to waste exploration time comparing suboptimal arms which are hard to differentiate from each other, and it can take a long time before the single good arm is identified. 


\captionsetup[subfloat]{labelformat=empty}


\subsubsection{Experiments on Real Learning-to-Rank Datasets}
\label{sec:LargeScale}

\begin{figure}
  \centering
  \subfloat[][MSLR]{
  \includegraphics[width=0.35\textwidth]{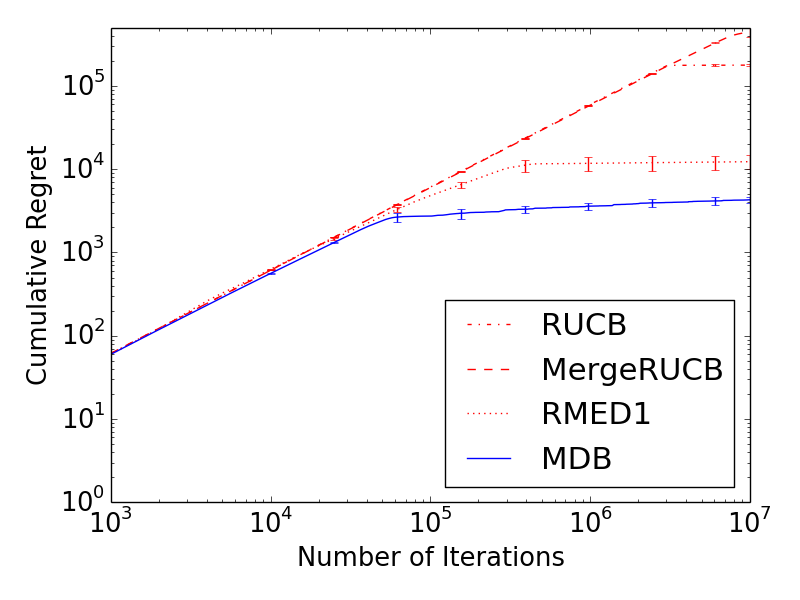}  
  } \\
  \subfloat[][YLR1]{
    \includegraphics[width=0.35\textwidth]{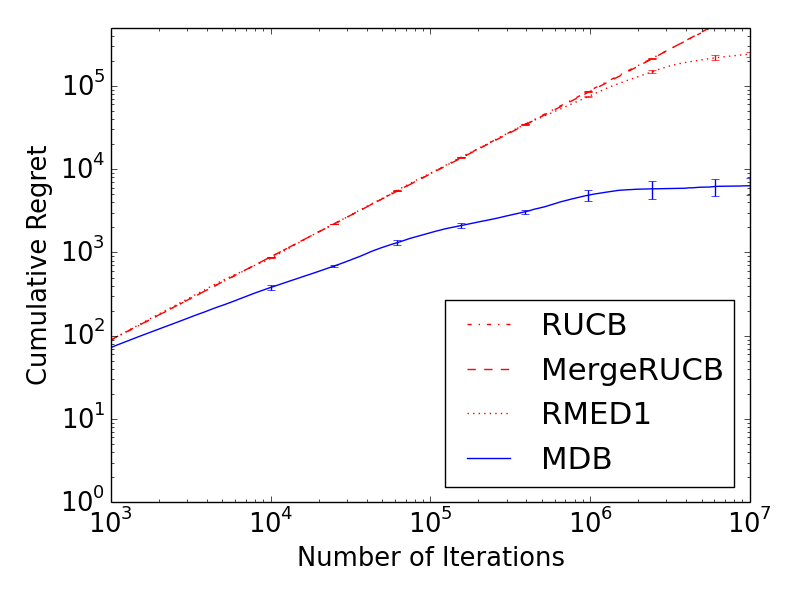}  
  } \\
  \subfloat[][YLR2]{
    \includegraphics[width=0.35\textwidth]{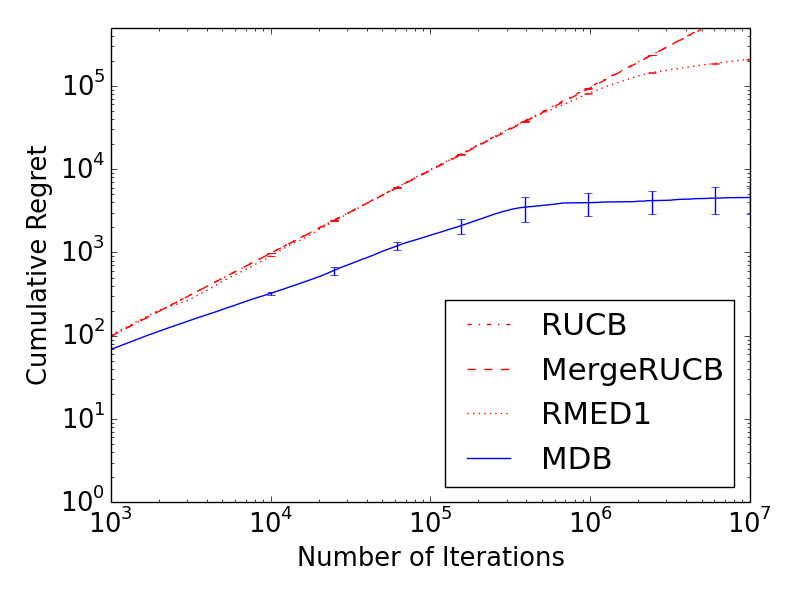}  
  } \\
  \subfloat[][Yandex]{
    \includegraphics[width=0.35\textwidth]{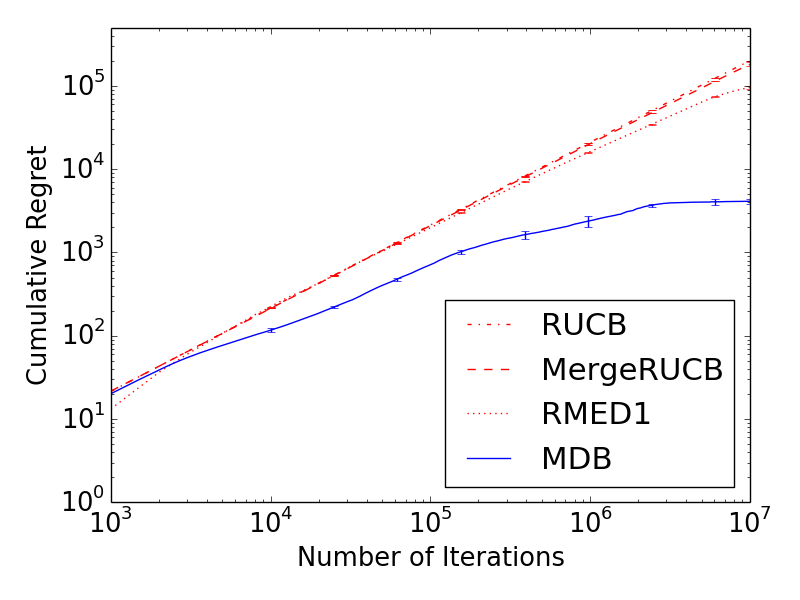}  
  }
\caption{Cumulative regret averaged over 10 runs against number of iterations for the 4 algorithms on the MSLR (a), YLR1 (b), YLR2 (c) and Yandex (d) datasets with the navigational click model. 
}
\label{fig:LargeScale}
\end{figure}

Figure~\ref{fig:LargeScale} shows how the cumulative regret increases with each iteration using the real data sets, MSLR, YLR1, YLR2 and Yandex. It clearly shows that MDB performs best for all the datasets. It outperforms the best dueling bandit algorithm, RMED1, by a factor of approximately 3 for the dataset with the smallest number of features, the MSLR dataset, and approximately 1-2 orders of magnitude for all the other datasets. RMED1 outperforms RUCB and MergeRUCB, as expected from the results of \cite{komiyama2015regret}. 


Note that for the Yandex dataset, since there was no Condorcet winner, we randomly sampled 200 of the 245 feature rankers to obtain a dataset with a Condorcet winner. Results using the full Yandex dataset with no Condorcet winner are described in Section~\ref{ss:noCondorcet}. 


\captionsetup[subfloat]{labelformat=parens}

\subsubsection{Dependence on Number of Rankers}

The results of Section~\ref{ss:synthetic} using synthetic data showed that the advantage of our algorithm increased relative to the three other algorithms, as the number of arms being compared increased. Additionally, the results of Section~\ref{sec:LargeScale} show that for the real datasets, the advantage of our algorithm ranges from a factor of approximately 3 for the MSLR dataset with the smallest number of features to approximately 2 orders of magnitude for the two YLR datasets, which have the greatest numbers of features. 


To isolate the impact of the number of rankers being compared on the real datasets involving multileaving, we investigate how regret scales with the number of rankers being compared using the YLR1 dataset. 
We randomly sampled subsets of rankers of sizes $\{10, 25, 40, 55, 70, 85, 100, 115, 130, 145\}$ from the YLR1 dataset. 
Note that we randomly sampled different subsets of rankers for each run. For each of these subsets we then carried out 10 runs of each algorithm over 5,000,000 iterations and recorded the average cumulative regret across runs. 

\begin{figure}
  \centering
  \includegraphics[width=0.9\columnwidth]{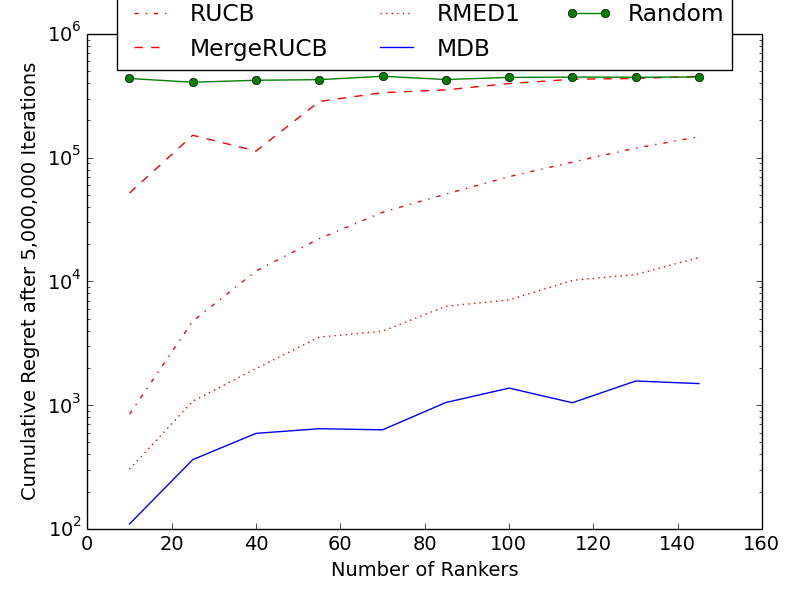}  
\caption{Cumulative regret averaged over 10 runs after 5,000,000 iterations against number of rankers for the 4 algorithms, and a random policy, on subsets with $M$ rankers of the YLR1 dataset with navigational click model. 
}
\label{fig:2}
\end{figure}

Figure~\ref{fig:2} shows how the performance of the 4 algorithms varies as a function of the number of rankers. Additionally we have shown the performance of a random policy which simply selects a random subset of the rankers for multileaving at each iteration.  We observe that as the number of rankers increases the cumulative regret increases most for RUCB and MergeRUCB, while it increases more slowly for RMED1. Regret appears to be almost independent of the number of rankers for MDB.

These experiments were also carried out for the perfect and informational click models, and the results were very similar but have been omitted due to page restrictions. 

For the MDB algorithm, the regret associated with having to explore suboptimal rankers does not appear to be additionally compounded by the number of rankers being explored. This is an important characteristic of the MDB algorithm, since if we can explore additional rankers with no substantial additional cost, the risks associated with large-scale online ranker evaluation are substantially mitigated. 


Note that it may appear that regret levels off for MergeRUCB as we increase the number of rankers. This is due to the fact that for 5,000,000 iterations there is a limit to how much regret can be incurred just by making random choices in 5,000,000 iterations. For larger problem sizes and for a time frame of 5,000,000 iterations, MergeRUCB begins to perform no better than a random policy. This does not imply that MergeRUCB performs as badly as a random policy in general, but for these problem instances it has not yet begun to eliminate suboptimal arms after 5,000,000 iterations. Further iterations would be needed to show improvements relative to the random policy. 


\captionsetup[subfloat]{labelformat=empty}

\subsubsection{Dependence on Click Model}
\label{sec:CM}

\begin{figure}
  \centering
  \subfloat[][perfect click model]{
  \includegraphics[width=0.45\textwidth]{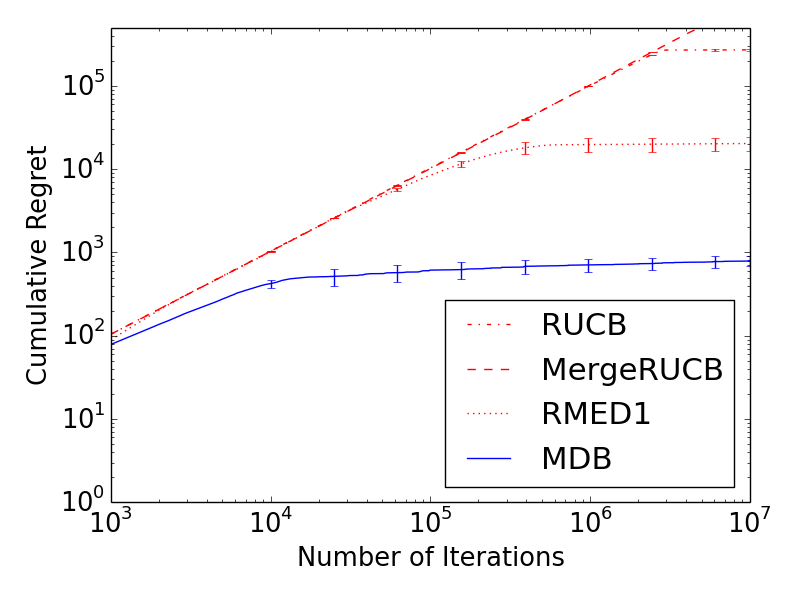}  
  } \\
  \subfloat[][navigational click model]{
    \includegraphics[width=0.45\textwidth]{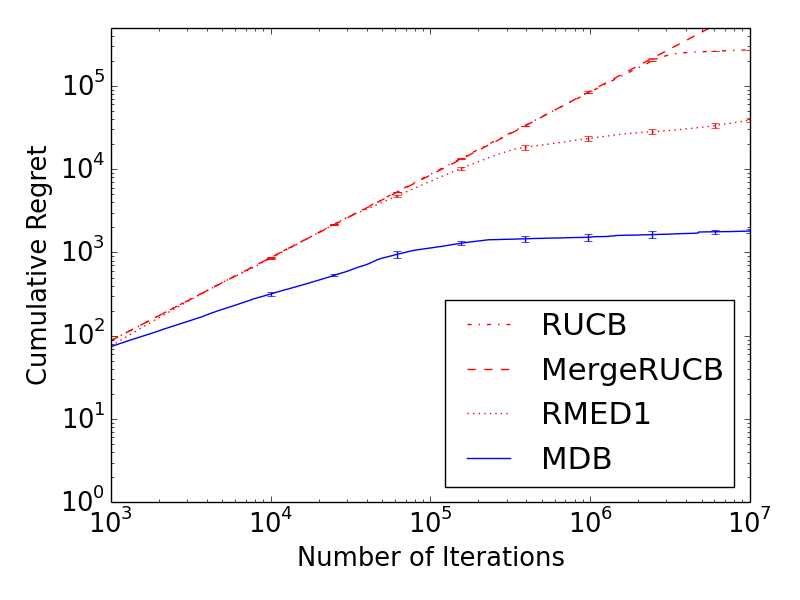}  
  } \\
  \subfloat[][informational click model]{
    \includegraphics[width=0.45\textwidth]{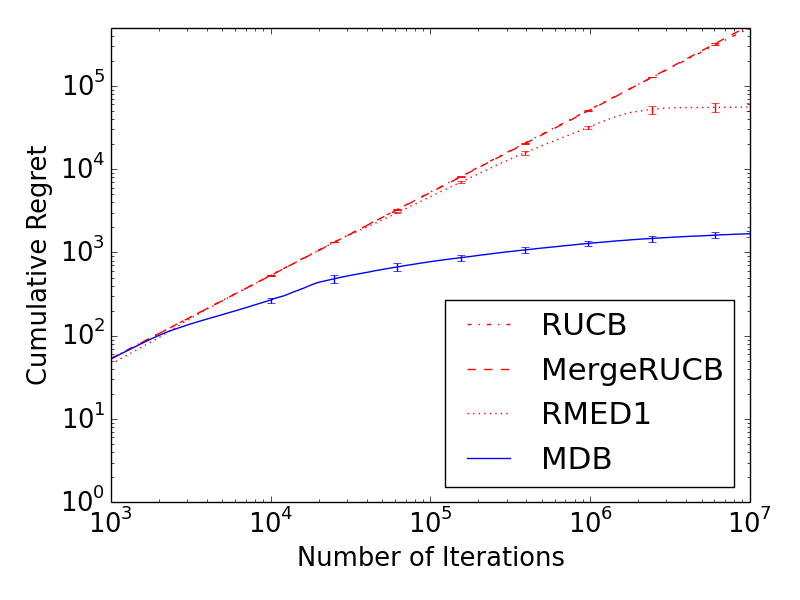}  
  }
\caption{Cumulative regret averaged over 10 runs against number of iterations for the 4 algorithms on the YLR1 dataset using the (a) perfect, (b) navigational and (c) informational click models. 
}
\label{fig:ClickModels}
\end{figure}

To test the robustness of our approach to the choice of click model, we also investigated performance using the perfect and informational click models \cite{hofmann2013fidelity}. These click models have less, respectively more, noisy user behaviour than the navigational model. Since different click models can reflect different types of user behaviour and search intent it is important that the algorithms are robust to different click models. Figure~\ref{fig:ClickModels} shows how the cumulative regret is affected by different user click models, using randomly selected subsets of size 200 of the rankers from the YLR1 dataset. We chose to use subsets of the full dataset for these experiments because of the computational costs of running RMED1 on the full YLR1 dataset. For all click models MDB outperforms the best dueling bandit algorithm by between 1 and 2 orders of magnitude. 

For MDB, the regret doubles when going from the perfect to the navigational click model, but does not increase further for the informational click model. In contrast, for the dueling bandit algorithms, regret for the informational click model is approximately double that for the navigational click model, which is approximately double that of the perfect click model. MDB is therefore least affected by varying the click model in our experiments. 


\captionsetup[subfloat]{labelformat=parens}






\subsubsection{Dataset without Condorcet winner}
\label{ss:noCondorcet}

As noted earlier, the baseline dueling bandit algorithms and our algorithm MDB assume the existence of a Condorcet winner, i.e. a ranker which beats every other ranker in expectation. In practice, this may not be true, and in fact there is no Condorcet winner for the full Yandex dataset.  To evaluate how the algorithms perform when the Condorcet assumption is violated, we investigated the performance of the algorithms on the full Yandex dataset. Since there is no Condorcet winner we cannot use the regret definition from Equation~\ref{eq:regret2} to evaluate the algorithms. Instead we define the winner for the full Yandex dataset based on the NDCG@10 score, denote this score by $NDCG^*$, and use a definition of regret given by 
\begin{equation}
r(S_t) = \frac{\sum_{j \in S_t}NDCG^*-NDCG_j}{|S_t|}.
\label{eq:regretNDCG}
\end{equation}

\begin{figure}
  \centering
  \includegraphics[width=0.9\columnwidth]{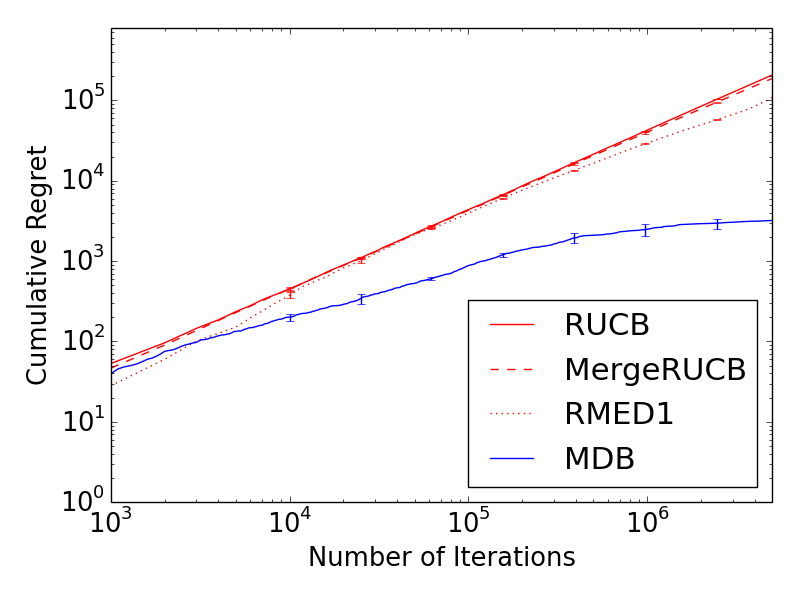}
\caption{Average cumulative regret over 10 runs against number of iterations for the 4 algorithms on the full Yandex dataset. 
}
\label{fig:noCondorcet}
\end{figure}

We carried out 10 runs of each algorithm over 5,000,000 iterations and recorded the average regret over runs at each iteration. Figure~\ref{fig:noCondorcet} shows how the cumulative regret increases with each iteration for the full Yandex dataset with regret defined in Equation~\ref{eq:regretNDCG}. The results are very similar to those from Figure~\ref{fig:LargeScale} for the Yandex subset with a Condorcet winner. MDB outperforms the best dueling bandit algorithm, RMED1, by approximately an order of magnitude after 5,000,000 iterations. 

\captionsetup[subfloat]{labelformat=empty}

\subsubsection{Distortion of probability estimates due to multileaving}
\label{sec:expConsistency}

As discussed earlier, simultaneous comparison of more than two arms may affect their pairwise winning probabilities. We called this effect {\em distortion}.  We can quantify this effect by first randomly sampling a fixed size subset of rankers that includes a Condorcet winner, and then measuring, after some fixed number of multileavings, the fraction of rankers that beat the Condorcet winner more than 50\% of the time. If there is no distortion, and the number of multileavings is sufficient, we expect this fraction to be zero.


In these experiments we test the level of distortion in the multileaving method SOSM, and examine how robust our MDB algorithm is to possible distortions in the multileaving method. 

For each dataset, and each click model, we randomly sample subsets of rankers of sizes 3, 10, and 100 that include a Condorcet winner. We examine the probabilities of the rankers beating the Condorcet winner after 3,000 multileavings. Note that this is likely to be an overestimate of the distortion of the multileaving method, since, for rankers of very similar quality, 3,000 iterations may not be sufficient to reliably distinguish rankers. Table~\ref{tab:inconsistency} shows the average, over 30 runs, of the percentage of rankers that beat the Condorcet winner.

We observe that the distortion problem is almost unique to the MSLR dataset, and is exacerbated by the noisier click models. 
The distortion problem is exclusively related to the feature ranker 133 in the MSLR dataset. Feature ranker 133 scores documents solely based on the query-document clicks, i.e. a document was clicked on in response to a query. This feature is very good at identifying 1 or 2 documents that are very likely to be relevant. However, when asked to rank documents in a multileaved set, most of the documents, even though they might be relevant, have not been previously clicked on. As such, ranker 133 is unable to distinguish between the vast majority of documents. Thus, even though ranker 133 performs well in pairwise comparisons, where it has contributed half of the documents in the results list, it performs very poorly when multileaved with many other rankers. 
Table~\ref{tab:inconsistency} also includes results for the MSLR dataset, when feature ranker 133 is excluded. This is denoted by MSLR*. When ranker 133 is excluded, no substantial distortion is observed.


The moderate levels of distortion observed for the Yandex and MSLR dataset (excluding feature ranker 133) are likely to be mostly caused by the fact that there are many rankers that are very similar in quality, and so 3,000 comparisons are not sufficient to differentiate these similar rankers. 

\begin{table}[ht]
\caption{Percentage of rankers beating the Condorcet winner (distortion), averaged over 30 runs, after 3,000 iterations for 3, 10, and 100 rankers being multileaved for each dataset and click model. The dataset denoted MSLR* is the MSLR dataset with feature ranker 133 removed.}
\begin{center}
\begin{tabular}{ | c | | c | | c | | c | }
  \hline
   & \multicolumn{3}{|c|}{Distortion}\\ \hline \hline
   Num. Rankers & 3  & 10  & 100  \\ \hline \hline
   MSLR Perfect & 0.0\% & 0.0\% & 9.2\% \\ 
   MSLR Navigational & 0.0\% & 0.0\% & 15.2\% \\ 
   MSLR Informational & 0.0\% & 6.8\% & 41.3\% \\ \hline
   MSLR* Perfect & 1.7\% & 3.1\% & 3.3\% \\ 
   MSLR* Navigational & 1.7\% & 4.0\% & 3.5\% \\ 
   MSLR* Informational & 0.0\% & 2.9\% & 2.7\% \\ \hline
   YLR1 Perfect & 0.0\% & 0.0\% & 0.0\% \\ 
   YLR1 Navigational & 0.0\% & 0.0\% & 0.0\% \\ 
   YLR1 Informational & 0.0\% & 0.0\% & 0.3\% \\ \hline
   YLR2 Perfect & 0.0\% & 0.0\% & 0.4\% \\ 
   YLR2 Navigational & 0.0\% & 0.4\% & 0.8\% \\ 
   YLR2 Informational & 0.0\% & 1.0\% & 0.9\% \\ \hline
   Yandex Perfect & 0.0\% & 1.1\% & 1.4\% \\ 
   Yandex Navigational & 3.3\% & 4.5\% & 3.8\% \\ 
   Yandex Informational & 3.3\% & 3.9\% & 3.4\% \\ \hline
\end{tabular}
\label{tab:inconsistency}
\end{center}
\end{table}

\begin{figure}
  \centering
  \includegraphics[width=0.9\columnwidth]{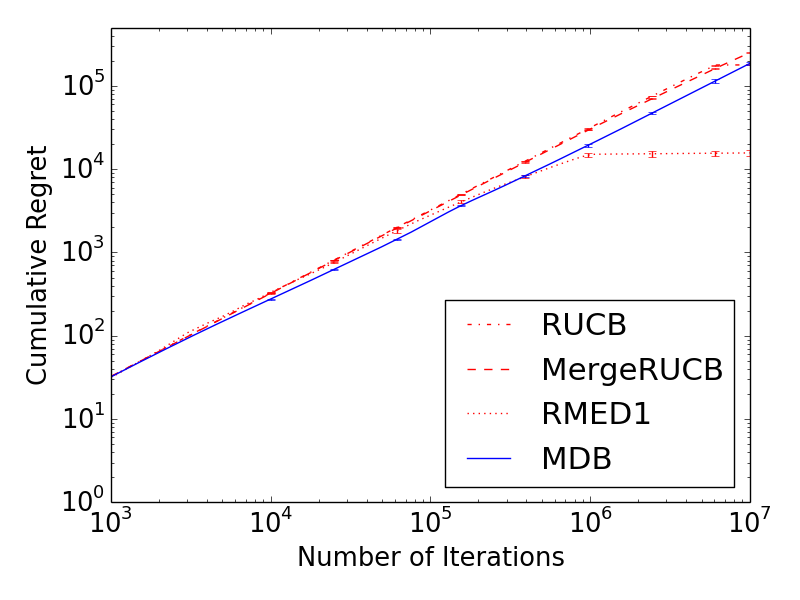}  
\caption{Average cumulative regret over 10 runs against number of iterations for the 4 algorithms on the MSLR dataset using the informational click model. 
}
\label{fig:MSLRInformational}
\end{figure}

The only problem setting where our MDB algorithm did not substantially outperform the best baseline dueling bandit algorithm, RMED1, was for the MSLR dataset with all 136 feature rankers with the informational click model. The results for this problem setting are shown in Figure~\ref{fig:MSLRInformational}. This is due specifically to the feature ranker 133 in the MSLR dataset. Table~\ref{tab:inconsistency} shows that there was some distortion for all click models for the MSLR dataset. However, it is with the informational click model that the distortion is greatest, reaching 41.3\% for 100 rankers. This is a very high percentage. The MDB algorithm appears to be robust to more reasonable levels of distortion,  suffering substantially less regret than the baselines for the MSLR dataset with the navigational click model, shown in Figure~\ref{fig:LargeScale}(a), and with the perfect click model (this result is omitted due to the page restriction). Additionally, for the MSLR dataset with feature ranker 133 removed, MDB substantially outperformed all baselines for all click models. Results are omitted due to the page restriction. 


\captionsetup[subfloat]{labelformat=empty}

\section{Conclusions and Future Work}

This paper proposes a generalisation of the $K$-armed dueling bandits 
termed multi-dueling bandits (MDB). 
%
%
We have applied MDB in online ranker evaluation to leverage the power of simultaneous comparisons through multileaving and improve the exploration-exploitation trade-off. Our experimental results on
synthetic data and real data from 4 standard datasets demonstrated up to 1 to 2 orders of magnitude reduction in regret compared to state-of-the-art dueling bandit algorithms, RUCB \cite{zoghi2014relative}, MergeRUCB \cite{zoghi2015mergerucb}, and RMED1 \cite{komiyama2015regret} in all except one pathological case discussed below.
Generally, relative benefits compared to dueling bandits increased with the number of rankers being compared. For MDB, the incurred regret did not increase substantially as the number of rankers increased. As such, the risks associated with large-scale online ranker evaluation are substantially mitigated. Further experiments showed that MDB was robust to various user click models.

Experiments were also conducted to examine the behaviour of MDB in the absence of a Condorcet winner, which is the case for the full Yandex dataset. In this case, the regret 
was approximated by measuring 
the 
NDGC@10
score. In this case MDB outperformed the best dueling bandit algorithm, RMED1, by approximately an order of magnitude after 5,000,000 iterations.


We also investigated 
the level of distortion of pairwise winning probabilities in multileaving using SOSM.
For the MSLR dataset using a navigational click model, the distortion reached 41.3\%. In this case MDB 
was inferior to RMED1. 
The high level of distortion was due to the peculiarities of ranker 133. 
If ranker 133 is removed, the distortion 
of pairwise winning probabilities is 
significantly reduced and MDB 
outperforms all other algorithms.

There are a number of avenues for future work. The distortion 
of pairwise winning probabilities in multileaving 
needs further investigation. All 
existing multileaving algorithms exhibit this behaviour. It remains an open question as to whether a new multileaving algorithm can be designed to avoid this problem, or at least minimize it. Furthermore, a theoretical analysis of our algorithm needs to be developed to better understand its power and limitations. 
Additionally, since a Condorcet winner is not guaranteed to exist, it may be useful to explore other concepts of winners, such as the Copeland \cite{zoghi2015copeland}, Borda \cite{urvoy2013generic} and von Neumann \cite{dudik2015contextual} criteria. Finally, we note that the proposed multi-dueling bandit algorithm can be applied to 
a broad class of problems and applications in other domains, e.g. recommender systems.

\bibliographystyle{abbrv}

\bibliography{Bib}

\end{document}